\documentclass[letter,11pt]{JHEP3}

\usepackage{amsmath,latexsym,amssymb}
\usepackage{psfrag}
\usepackage[latin1]{inputenc}
\usepackage{subfigure}
\usepackage{nicefrac}
\usepackage{bbm}
\usepackage[stable]{footmisc}
\usepackage{fancyheadings}
\usepackage[pdftex]{graphicx}
\usepackage{psfrag}
\usepackage{fancybox}
\usepackage{ifthen}
\usepackage{epsfig}
\usepackage[errorshow]{tracefnt}
\usepackage{multirow}
\newcommand\bbone{\ensuremath{\mathbbm{1}}}

\newcommand\beal{\begin{align}}

\renewcommand{\theenumii}{\alph{enumii}}

\newcommand\nn{\nonumber}

\newcommand{\eq}[1]{\begin{equation}#1\end{equation}}

\newcommand{\spl}[1]{\begin{split}#1\end{split}}

\def\Re           {{\rm Re\hskip0.1em}}
\def\Im           {{\rm Im\hskip0.1em}}

\newcommand\benu{\begin{enumerate}}
\newcommand\eenu{\end{enumerate}}
\newcommand\bit{\begin{itemize}}
\newcommand\eit{\end{itemize}}

\newcommand\cm{\mathcal{M}}

\newcommand\ie{{\it i.e.}}

\newcommand\nr{N_{\mathbb{R}}}

\newcommand{\boxedeq}[1]{
\begin{equation}
\fbox{
\rule[0.7cm]{0pt}{0pt}
$#1$
\rule[-0.45cm]{0pt}{0pt}
}
\end{equation}
}

\title{Flux compactification on smooth, compact three-dimensional 
toric varieties}

\author{Magdalena Larfors${}^{\clubsuit}$, Dieter L\"{u}st${}^{\diamondsuit\clubsuit}$
 and Dimitrios Tsimpis${}^{\clubsuit}$ \\

  \begin{itemize}

\item  Max-Planck-Institut f\"ur Physik\\
F\"ohringer Ring 6, 80805 M\"unchen, Germany
  
\item  Arnold-Sommerfeld-Center for Theoretical Physics\\
Department f\"{u}r Physik, Ludwig-Maximilians-Universit\"{a}t M\"{u}nchen\\
Theresienstra\ss e 37, 80333 M\"{u}nchen, Germany
  \end{itemize}

\bigskip
 E-mail:
\email{magdalena.larfors@physik.uni-muenchen.de}, \email{dieter.luest@lmu.de} \& \email{luest@mppmu.mpg.de}, \email{dimitrios.tsimpis@lmu.de} }

\abstract{Three-dimensional smooth, compact 
toric varieties (SCTV), when viewed as real six-dimensional manifolds, 
can admit $G$-structures rendering them suitable for internal manifolds in  supersymmetric flux compactifications. 
We develop techniques which allow us to systematically construct $G$-structures  
on SCTV and read off their 
torsion classes. We illustrate our methods with explicit examples, one of which 
consists of an infinite class of toric $\mathbb{CP}^1$ bundles. We give a self-contained review of the relevant concepts from toric geometry, 
in particular  the subject of the classification of SCTV in dimensions $\leq3$. 
Our results open up the possibility for a systematic construction and study of supersymmetric flux vacua based on SCTV.
}

\keywords{Flux compactifications, Toric varieties}
\preprint{LMU-ASC 30/10\\MPP-2010-54}
\begin{document}

\section{Introduction}

In the absence of background fluxes and sources, supersymmetric compactifications of string theory in the large-volume limit are of the form ${\mathbb R}^{3,1}\times \mathcal{M}_6 $, where $\mathcal{M}_6$ is a  Calabi-Yau (CY) three-fold. Turning on fluxes  generally forces the internal manifold $\mathcal{M}_6$ to no longer be CY. 
Despite a great amount of recent progress on the subject of flux compactifications, the geometry of $\mathcal{M}_6$ is in general 
not as well understood as in the case without fluxes. This largely stems from the fact that as soon as 
fluxes are turned on, the internal manifold ceases to be complex in general\footnote{There are well-known exceptions, e.g. the warped 
CY vacua of IIB.} and 
the powerful methods of algebraic geometry are no longer available. 

Necessary, but not sufficient, conditions on the geometry of  $\mathcal{M}_6$ in the presence of fluxes can be stated in the language of generalized geometry \cite{hitchin, gualtieri, gran}. Necessary and sufficient conditions exist for supersymmetric $SU(3)$-backgrounds of IIA  \cite{lt}, which  can be thought of as a generalization of CY backgrounds: the latter are obtained from the former in the limit of vanishing Romans mass.\footnote{By that we mean a limit 
of the equations of type II supergravity; not a limit of any particular solution thereof. Moreover this is not a continuous limit, once flux quantization is taken into account.} Once fluxes are turned on, supersymmetry `selects' 
a generally non-integrable almost complex structure on $\mathcal{M}_6$.

There exist tens of thousands of CY's giving rise to equally many ten-dimensional vacua of the form ${\mathbb R}^{3,1}\times \mathrm{CY}_3$ (each of which typically comes with many moduli). On the other hand, only four classes of vacua (typically without any continuous parameters) of the form of \cite{lt} are known so far. In all of them the internal manifold admits a coset description \cite{klt},\footnote
{For early work on superstring compactification on coset spaces with $H$-flux and torsion see
\cite{Lust:1986ix}.}  and the 
supersymmetric solution is expressed in terms of the non-integrable almost complex structure on $\cm_6$. However, 
as pointed out in \cite{toma}, two of these cosets, $Sp(2)/S(U(2)\times U(1))$ and $SU(3)/U(1)\times U(1)$, can be thought of as twistor spaces and therefore also admit a complex structure.\footnote{Interestingly, the twistor space description of the coset $SU(3)/U(1)\times U(1)$ does not cover the entire three-dimensional parameter space, but only a two dimensional subspace thereof \cite{klt}.}  

The above observation hints at the possibility that we may have our cake 
and eat it too: by taking $\cm_6$ to be a complex manifold, an algebraic  
variety in particular, we may still make use of the toolkit of algebraic 
geometry -- provided of course  $\cm_6$ also admits a second non-integrable 
almost complex structure in accordance with the general conditions  for 
a supersymmetric vacuum in the presence of fluxes. The class of three-dimensional (six real dimensions) smooth, compact toric varieties (SCTV) is a natural candidate.

Because there are no compact toric CY manifolds, three-dimensional compact toric varieties have not received much attention in the physics literature.  In using toric geometry to construct compact three-dimensional CY's, one  instead looks for four- or higher-dimensional toric varieties in which the CY is embedded as a submanifold. Thereby, toric geometry has proved to be a very useful tool for the construction of CY-spaces (see   \cite{Kreuzer:2002uu}  for a toric computational  package of CY-spaces). Nevertheless, three-dimensional SCTV do not  play any  role in this construction. As a consequence the main result on the subject -- the (partial) classification of three-dimensional SCTV \cite{ota} -- remains, to our knowledge, rather obscure.

Here we show that in the presence of flux the situation is changed drastically: the internal manifold is no longer CY and the possibility arises that $\mathcal{M}_6$ can be taken to be a  SCTV. Indeed this is the case for the celebrated $AdS_4\times \mathbb{CP}^3$ example \cite{np} (which is a special point in the moduli space of the coset example $Sp(2)/S(U(2)\times U(1))$ of \cite{toma,klt}): as is well-known, in any dimension the complex projective space admits a toric description. 
The presence of fluxes therefore opens up the possibilty to draw upon
 the vast `playground' of SCTV  in looking for internal manifolds in flux  compactifications. For the reasons just explained, this has never been considered before.

Toric varieties are usually described in the 
mathematics literature in terms of { fans}, which are 
collections of cones obeying certain defining properties, so that for 
every such fan $\Sigma$ there is a unique toric variety $V_{\Sigma}=\cm_6$ 
associated with it ({\it cf.} the note on notation at the end of this section). 
The advantage of this approach is that there exists 
a simple dictionary between many of the 
topological or analytic properties of $V_{\Sigma}$ and the 
properties of  $\Sigma$. On the other hand, there exists  
an equivalent description of toric varieties 
as  symplectic quotients. This latter approach is perhaps 
more transparent from the physics point of view as it lends itself 
to a gauged linear sigma-model interpretation \cite{witten}.

Given the symplectic quotient description of a three-dimensional SCTV $V_\Sigma$, it is straightforward to construct an ansatz for an $SU(3)$ structure on $\cm_6=V_\Sigma$. However, the symplectic quotient description is given in terms of a certain set of $U(1)$ charges $Q^a$, whereas the classification of three-dimensional toric varieties is given in terms of certain weighted triangulations of the two-sphere \cite{ota}. We therefore need a translation between the two descriptions. 

In the present paper we will explain how to read off from a so-called `double $\mathbb{Z}$-weighted' two-sphere triangulation the set of fundamental generators $G(\Sigma)$ of the fan $\Sigma$ with which the toric variety $V_\Sigma$ is associated. Moreover we explain how to read off from $G(\Sigma)$ the charges $Q^a$ of $V_\Sigma$ in the symplectic quotient description; using the latter  we will be able to 
construct $SU(3)$-structures on $\mathcal{M}_6$ and  read off the corresponding torsion classes.

Our strategy is summarized 
below:

\begin{center}
{ {\bf The road from toric geometry to $G$-structures}}

\bigskip

{\small \begin{tabular*}{15cm}[t]{cccccc}
\framebox{\parbox[c]{3.8cm}{\begin{center}3D classification of $V_\Sigma$\\ (weighted triangulation)\end{center}}} &$\longrightarrow$ & 
\framebox{\parbox[c]{3.8cm}{\begin{center}fan $\Sigma$\\ (generators $G(\Sigma)$)\end{center}}} & $\longrightarrow$&\\
&&&&\\
 &$\longrightarrow$ 
& \framebox{\parbox{3.8cm}{\begin{center}symplectic quotient\\ ($U(1)$ charges $Q^a$)\end{center} }}&$\longrightarrow$ & \framebox{\parbox{3.8cm}{\begin{center} geometry of $\cm_6=V_\Sigma$\\
($SU(3)$ structure)  \end{center}}}& \\
\end{tabular*}}

\end{center}

\bigskip

The aim of this paper is to explain the above diagram in detail. 

The remainder of this paper is organized as follows: In section 
\ref{toricreview} we begin with a self-contained review of the 
relevant facts from toric geometry, with the main emphasis on the 
classification of  three-dimensional SCTV presented in 
section \ref{classification}. Since these results are less well-known within the 
physics community we have chosen to review them in some detail. 
An important subclass of three-dimensional 
SCTV consists of certain $\mathbb{CP}^1$ bundles over two-dimensional SCTV, hence the classification of two-dimensional SCTV is also reviewed in detail. 

The symplectic quotient description of toric varieties is reviewed 
in section \ref{symplectic}. It is this description that we will eventually 
use in order to construct $G$-structures on $\mathcal{M}_6$. With that in mind, 
we recast the classification results of section \ref{classification} in 
the language of symplectic quotients at the end of section \ref{symplectic}, where we also list the charges of the first few  toric 
$\mathbb{CP}^1$ bundles mentioned in the previous paragraph. 

\vfill\break

In section \ref{su3a} we give a brief review of $SU(3)$ structures on $\mathcal{M}_6$. In section \ref{sec:4.2} we give a simple ansatz for $SU(3)$ structures on $\mathcal{M}_6=V_\Sigma$  in terms 
of toric data of $V_\Sigma$. We illustrate our method with a couple of explicit examples in section \ref{examples}. We conclude with some discussion of future directions in section \ref{conclusions}.

{{\bf A note on notation:}} We will denote by $V_\Sigma$ a toric variety 
associated with a fan $\Sigma$.   The complex dimension of $V_\Sigma$ 
will be denoted by $d=\mathrm{dim}_{\mathbb{C}}V_\Sigma$. When viewed as a real $2d$-dimensional manifold, we will denote  $V_\Sigma$ by $\mathcal{M}_{2d}$. 
The number 
of fundamental generators $v_i\in G(\Sigma)$ of $\Sigma$ will be denoted by $n:=\mathrm{dim}G(\Sigma)$. We will denote by $s$ the difference $s:=n-d$; 
in the symplectic quotient description $s$ is the number of 
$U(1)$ charges $Q^a$.

\section{Review of toric geometry\label{toricreview}\footnote{A classic introductory text to the subject of toric geometry is \cite{toric}. 
For recent pedagogical introductions to toric geometry 
for physicists the reader may consult \cite{Reffert:2007im,denef}.}}

Toric varieties are usually defined in the 
mathematics literature in terms of { fans}, which are 
collections of cones obeying certain  properties 
to be discussed in more detail momentarily, so that for  
every fan $\Sigma$ there is a unique toric variety $V_{\Sigma}$ 
associated with it. The advantage of this approach is that there exists 
a simple dictionary, to be reviewed below,  between many of the 
topological, algebraic  or analytic properties of $V_{\Sigma}$ and the 
properties of  $\Sigma$. 

To set the stage, we will need a number of definitions. 
Consider a rank-$d$ integer lattice $N\cong \mathbb{Z}^d$. We shall 
denote by $N_{\mathbb{R}}$ the real extension of $N$, 
$N_{\mathbb{R}}:=\mathbb{R}\otimes N$. A subset $\sigma\subset \nr$ 
is a called a {\it strongly convex rational polyhedral cone with apex $0$} 
(simply a {\it cone} henceforth) if $\sigma\cap(-\sigma)=\{0\}$ and there 
exist elements $v_1,\dots, v_r$ of $N$  such that
\eq{
\sigma=\{ a_1v_1+\dots a_rv_r~; ~~~0\leq a_1,\dots a_r\in \mathbb{R} \}~.
}
In the following we will assume that $v_1,\dots, v_r$ are linearly independent 
in $\mathbb{R}$ and  {\it primitive} ( \ie{} for each $i=1,\dots, r$, $v_i$ is not a non-trivial integral multiple of any element of $N$). 
With the above assumptions,  $v_1,\dots, v_r$ are uniquely determined for a given $\sigma$ and are called the {\it fundamental generators of $\sigma$} 
(or simply {\it generators}); moreover $\sigma$ is a {\it simplicial cone}. 

\begin{figure}
\centering
\includegraphics[height=5cm,width=4cm, viewport=60 120 510 675,clip]{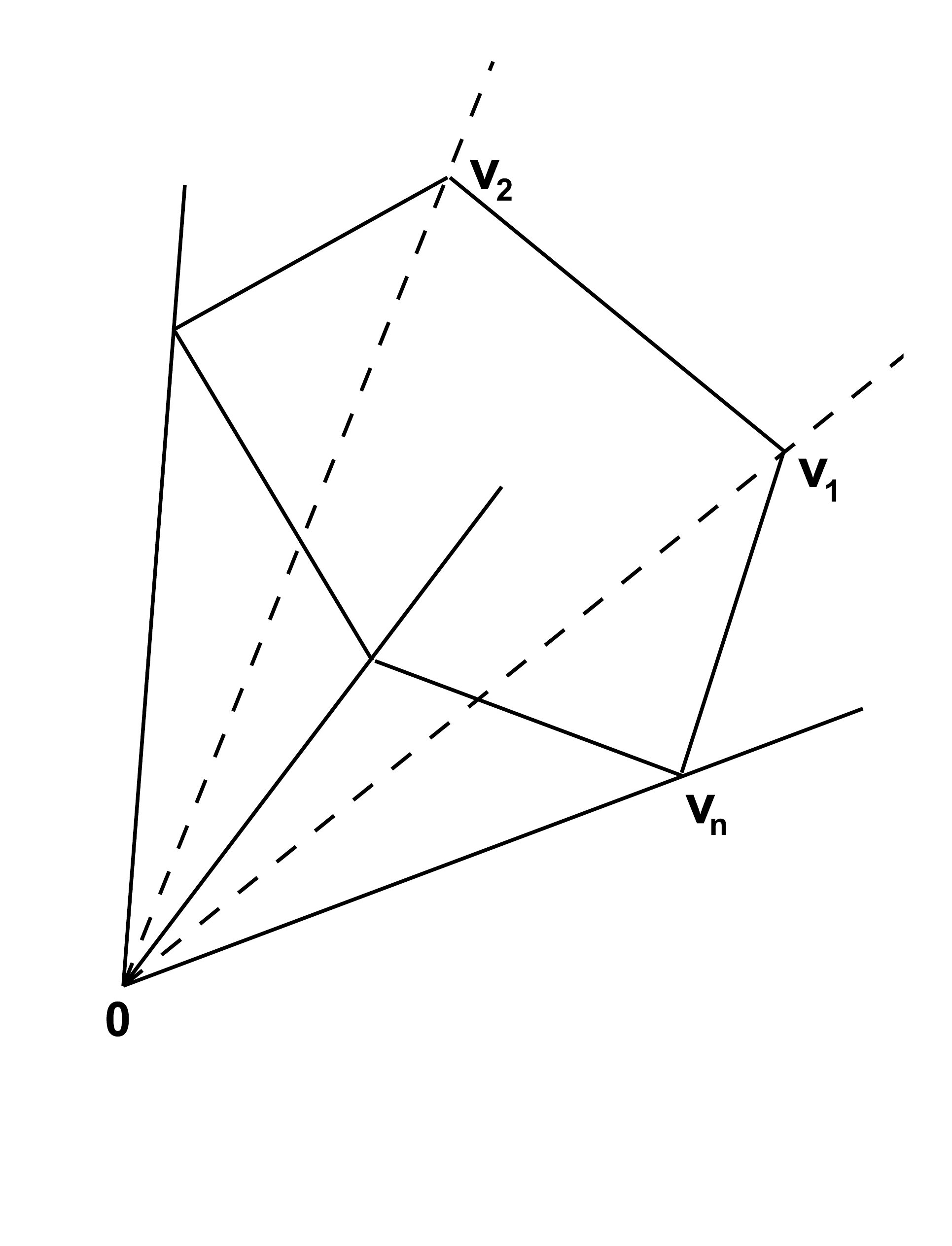}
\caption{A maximal cone in a three-dimensional space $\nr$ with  one-dimensional fundamental cone generators $(v_1, v_2, ..., v_n)$.}
\label{oda6}
\end{figure}

The {\it dimension of $\sigma$} 
is the dimension of the real vector subspace $\sigma+(-\sigma)$ in $\nr$. A 
cone $\sigma$ in $N$ is called {\it maximal} if its dimension is 
equal to the rank of $N$. We call $\sigma'$ a {\it face} of $\sigma$ (denoted  $\sigma'\leq\sigma$) if its 
fundamental generators are a subset of the fundamental generators of $\sigma$; 
we identify the origin of $N$ with the cone generated by $\varnothing$. If  $\mathrm{dim}~\!\sigma'=\mathrm{dim}~\!\sigma-1$, and $\sigma'\leq\sigma$, we call $\sigma'$ a {\it facet} of $\sigma$.

A set of cones as defined above 
$\Sigma=\{\sigma_1,\dots, \sigma_k\}$ in $\nr$ is a {\it simplicial fan} 
(a {\it fan} henceforth) if it satisfies the following conditions:
\begin{itemize}
\item if $\sigma\in\Sigma$ and $\sigma'\leq\sigma$ then $\sigma'\in\Sigma$;
\item if $\sigma'\in\Sigma$ then there exists $\sigma\in\Sigma$ such that  $\sigma'\leq\sigma$ and $\sigma$ is maximal; 
\item if $\sigma, ~\!\sigma'\in\Sigma$ then $\sigma\cap\sigma'\leq\sigma$ and 
$\sigma\cap\sigma'\leq\sigma'$.
\end{itemize} 
We will denote by $G(\Sigma)$ the set of all 
fundamental generators of cones in $\Sigma$. 

\begin{figure}
\centering
\includegraphics[height=7cm,width=6cm, viewport=15 50 530 675,clip]{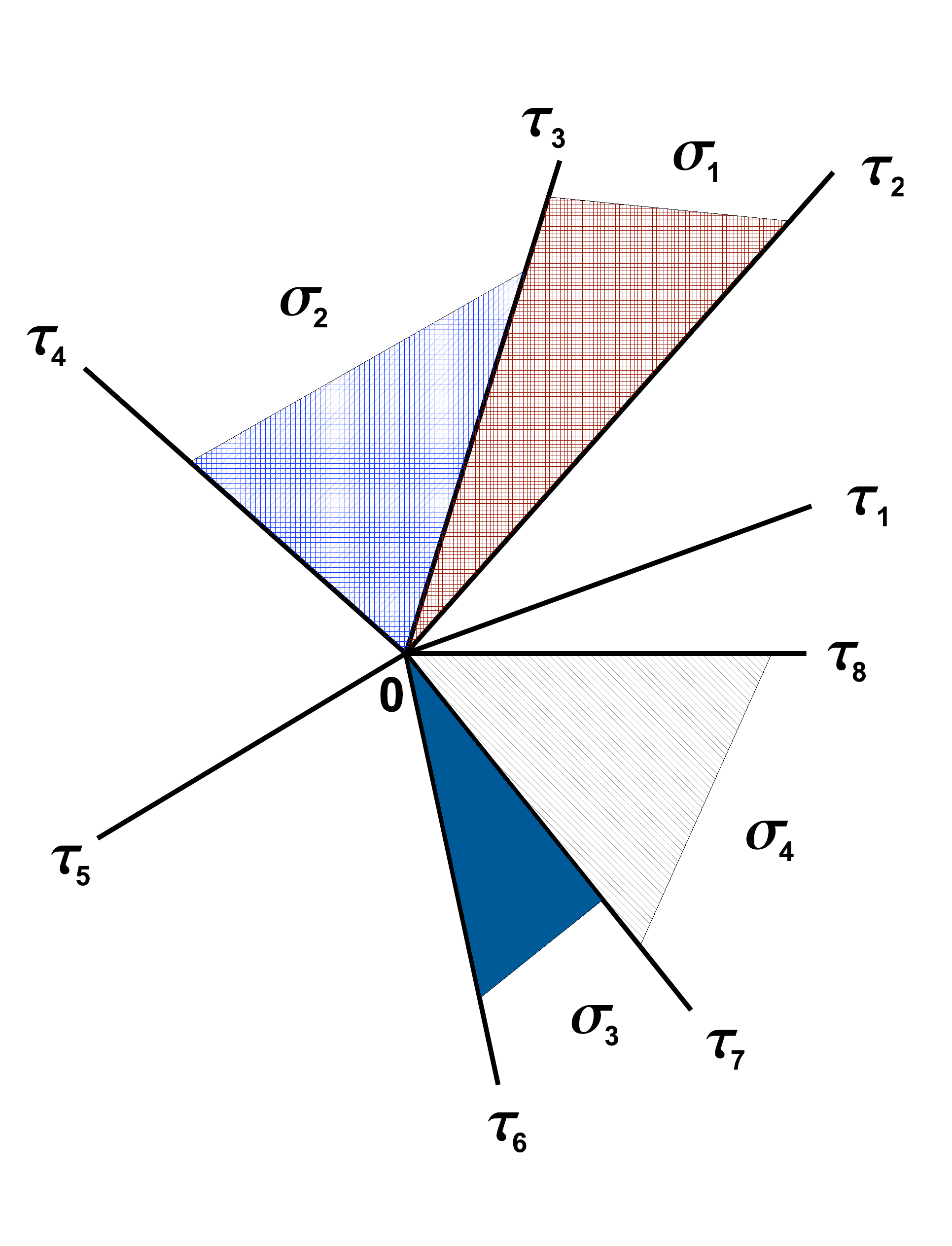}
\caption{A two dimensional fan $\Sigma$ consisting of four maximal cones ($\sigma_i$), eight one-dimensional cones ($\tau_i$) and one zero-dimensional one ($0$).}
\label{oda7}
\end{figure}

We are now ready to state the correspondence between the properties 
of toric varieties which are of relevance to our present discussion 
and the properties of the fan with which the variety is associated (this  
is summarized in table \ref{tab1}): 

(i) A toric variety  $V_{\Sigma}$ associated with a fan $\Sigma$ is {\it compact} (sometimes also called {\it complete} in the mathematics literature) if and only if $\nr$ is the union of cones in $\Sigma$. Fig. \ref{oda7} shows 
a two-dimensional fan  consisting of four maximal cones, eight one-dimensional cones and one zero-dimensional one: $\Sigma=\{\sigma_1,\dots, \sigma_4,\tau_1,\dots,\tau_8,0\}$. Since $\Sigma$ does not cover $\nr\cong\mathbb{R}^2$, the associated variety $V_\Sigma$ is not compact.

(ii) A toric variety  $V_{\Sigma}$ associated with a fan $\Sigma$ is 
{\it smooth} (also called {\it nonsingular} in the mathematics literature) if and only if 
a subset $\{v_1, \dots, v_d\}$ of $G(\Sigma)$ furnishes a $\mathbb{Z}$-basis of $N$ whenever  $\{v_1, \dots, v_d\}$  spans a maximal cone in $\Sigma$. 

(iii) Let a $P_{\Sigma}$ be a convex polytope with vertices in $N$, 
containing the origin of $N$ in its interior. A toric variety  $V_{\Sigma}$ associated with a fan $\Sigma$ is {\it projective} (\ie{} can be embedded in $\mathbb{CP}^r$ for some $r$) if and only if each cone in $\Sigma$ is spanned by the origin of $N$ and a face of $P_{\Sigma}$. 

(iv) In the previous correspondence, the vertices of 
$P_\Sigma$ need not be primitive vectors; if they are, the corresponding 
variety $V_\Sigma$ is called {\it toric Fano}. 
Note that sometimes a Fano variety is defined 
in the literature to have the additional property of being
 smooth; following \cite{kapr} we will here adopt the terminology {\it smooth Fano} for such varieties.

More generally, a {\it Fano} variety is a projective variety whose anticanonical divisor is ample. As follows from the previous discussion, isomorphism classes of $d$-dimensional toric Fano varieties are in one-to-one correspondence with isomorphism classes of convex lattice polytopes of dimension $d$ with a fixed lattice point inside (see \cite{borisov, kapr} for a review).

A smooth two-dimensional Fano variety is usually called a {\it del Pezzo surface}; exactly five of these are toric: $\mathbb{CP}^1\times\mathbb{CP}^1$, 
$\mathbb{CP}^2$, $\mathbb{CP}^2$ blown up at one, two and three generic points. The corresponding two-dimensional fans are shown in fig. \ref{2dfano}. In three complex dimensions there exists a complete classification of smooth toric Fano threefolds \cite{bat}: they correspond to 18 three-dimensional polytopes with a maximum of eight vertices.

\begin{table}[h!]
\begin{center}
\begin{tabular}{|c||c|}
\hline
$V_{\Sigma}$ & $\Sigma$ \\
\hline
\hline
compact & covers $N_{\mathbb{R}}$\\
\hline
smooth & $|\mathrm{det}(v_1,\dots, v_d)|=1$ for each $\{v_1, \dots, v_d\}$ spanning 
a maximal cone in $\Sigma$\\
\hline
projective & \parbox[t]{9.9cm}{each $\sigma\in\Sigma$ is spanned by the origin $O$ of $N$ and a face of a  convex polytope $P_{\Sigma}$ with $O$ in its interior and vertices in $N$} \\
\hline 
Fano & as above  and the vertices of $P_{\Sigma}$ are primitive 
elements of $N$\\
\hline 
\end{tabular}
\caption{Correspondence between properties of the fan $\Sigma$ and the associated toric variety $V_{\Sigma}$}
\label{tab1}
\end{center}
\end{table}

\subsection{Classifications}\label{classification}

SCTV with up to eight generators have been classified 
in \cite{ota}, while 
all  $d$-dimensional SCTV with $n=d+2$ generators have been classified in \cite{klein}. In the symplectic-quotient description, the number of $U(1)$ charges $(s)$ is equal to the number of generators $(n)$ minus the complex dimension of $V_{\Sigma}$: $d=n-s$, as will be reviewed in section \ref{symplectic}. In particular, note that $d$-dimensional toric varieties with $n=d+1$ generators are weighted projective spaces (for a  review of some of the properties of the latter and original references see section 5 of \cite{hubs}).

In \cite{ota},  the 
 classification of SCTV in two and three (complex) 
dimensions was shown to reduce to the classification of certain 
weighted circular graphs and weighted triangulations of the two-dimensional 
sphere, respectively. In the following we shall review these classifications. 
Ultimately we are interested in a translation of the results of \cite{ota} 
into the symplectic quotient language reviewed in section \ref{symplectic}; for that we will need to explain the classification results of \cite{ota} in some detail.

We will make use of the following:

{\it Lemma}. Consider a smooth, compact $d$-dimensional toric variety $V_\Sigma$ and a facet $\tau\in\Sigma$. Let $\{v_1,\dots,v_{d-1}\}$ be the fundamental 
generators of $\tau$. Then there exist exactly two maximal cones $\sigma$, $\sigma'\in\Sigma$ such that $\tau\leq\sigma$ and $\tau\leq\sigma'$. If 
$v$, $v'$ are the additional fundamental generators of $\sigma$, $\sigma'$, then 
there exist integers $a_i$, $i=1,\dots,d-1$, such that:
\eq{\label{central}
v+v'+\sum_{i=1}^{d-1}a_iv_i=0
~.}
Indeed, the existence of $\sigma$, $\sigma'$ follows from the fact that $\nr$ is the union of cones in $\Sigma$, since 
$V_\Sigma$ is compact. Equation (\ref{central}) follows from the fact that the fundamental generators of $\sigma$ and 
$\sigma'$, $\{v_1,\dots,v_{d-1}, v\}$ and $\{v_1,\dots,v_{d-1}, v'\}$, furnish two different $\mathbb{Z}$-bases of $N$, since
$V_\Sigma$ is smooth.

\subsection*{Two-dimensional classification}

\begin{figure}
\centering
\includegraphics[height=3.8cm,width=9.7cm,viewport=25 315 600 540,clip]{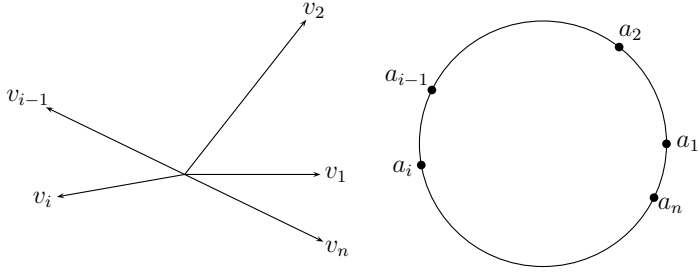}
\caption{A two dimensional toric variety $V_\Sigma$ determined by  primitive vectors $v_i\in N$, $i=1,\dots, n$, and its associated weighted circular graph.}
\label{Oda50}
\end{figure}

Let $d=2$ and $N\cong\mathbb{Z}^2$. A two-dimensional 
SCTV $V_\Sigma$ is determined by a set of 
 primitive vectors $v_i\in N$, $i=1,\dots, n$, going counterclockwise around the origin of $N$ in this order, such that $|\mathrm{det}(v_i,v_j)|=1$ for $i\neq j$ and $G(\Sigma)=\{v_1,\dots,v_n\}$. Form eqn.~(\ref{central}) it then follows that there exist 
integers $a_i$ such that
\eq{ \label{central2}
v_{i-1}+v_{i+1}+a_iv_i=0~,~~~0\leq i\leq n ~,
}
and we have set $v_0:=v_n$, $v_{n+1}:=v_1$. We can therefore  associate 
with $G(\Sigma)$ a {\it circular graph with weights $a_i$} (see fig.~\ref{Oda50}).

\begin{figure}
\centering
\includegraphics[height=10.7cm,width=12cm,viewport=20 240 570 705,clip]{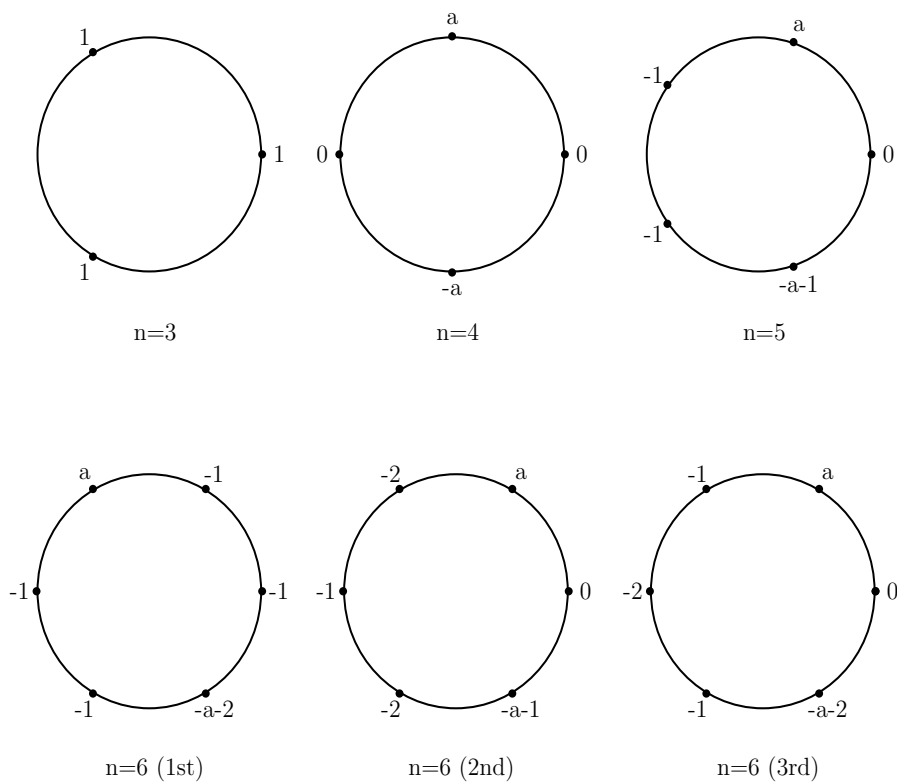}
\caption{Two-dimensional admissible weighted circular graphs with number 
of vertices $n\leq6$.}
\label{circular}
\end{figure}

Let us define an {\it admissible weighted circular graph} to be a weighted circular graph which is obtained from either of the first two circular graphs shown in fig.~\ref{circular} (those with $n=3,4$) by successive application of the following operation: insert a vertex with 
weight $-1$ and subtract $1$ from the weights of the two adjacent vertices. Note that performing this operation on the first graph of fig.~\ref{circular}  (the one with $n=3$), one obtains the second  circular graph (the one with $n=4$) for the special case $a=1$. Hence all admissible graphs with $n\geq 5$ are obtainable from the one with $n=4$.

The significance of the above is that the classification of two-dimensional toric varieties can be reduced to the classification of admissible circular graphs \cite{ota}: 

\begin{center}
\framebox{\parbox[t]{14.5cm}{\it
A weighted circular graph corresponds to a two-dimensional compact, smooth 
toric variety if and only if it is admissible.  
}}
\end{center}
Moreover, as shown in \cite{ota}, there is always some $i$ such that $v_n+v_i=0$, and we may take $\{v_1,v_i\}$ to be a $\mathbb{Z}$-basis of $N$. Taking  this into account, from an admissible weighted circular 
graph we can read off the corresponding $G(\Sigma)=\{v_1,\dots,v_n\}$  by solving eqs.~(\ref{central2}).

\begin{figure}
\centering
\includegraphics[height=14cm,width=12cm,viewport=40 160 570 725,clip]{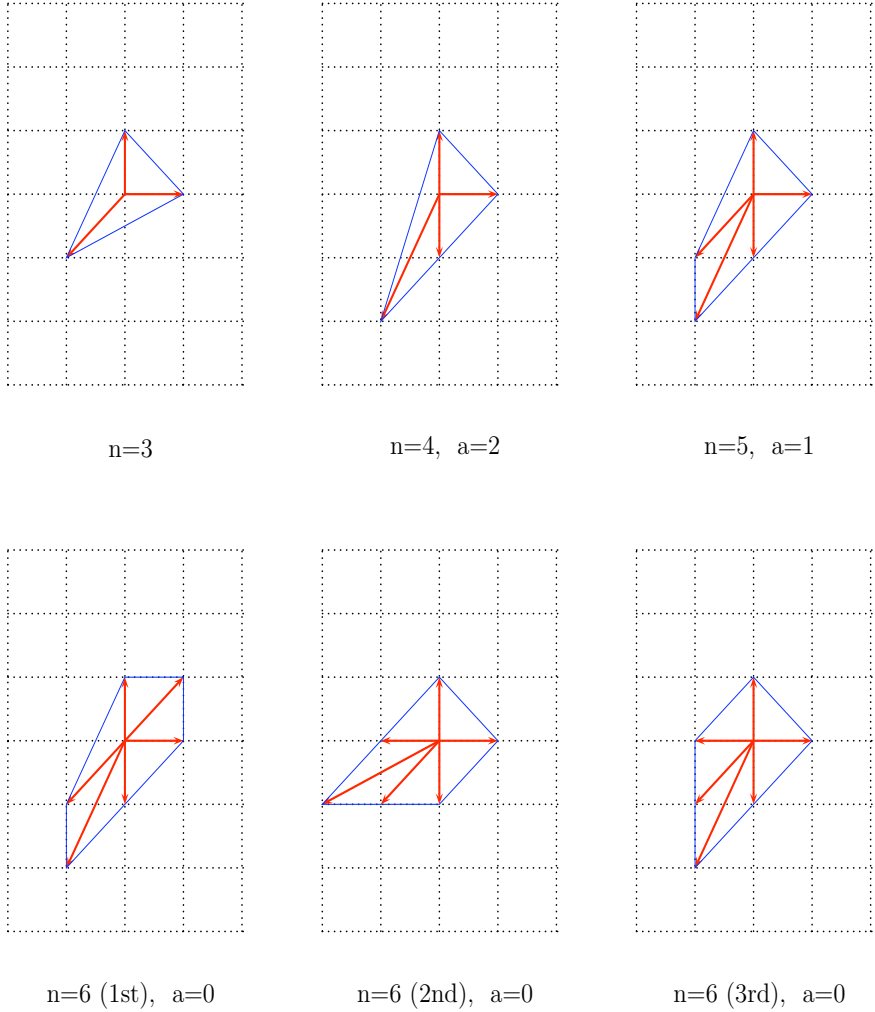}
\caption{Two-dimensional fans $\Sigma$ and fundamental 
generators corresponding to admissible weighted circular graphs  with number 
of vertices $n\leq6$.}
\label{2d}
\end{figure}

\begin{figure}
\centering
\includegraphics[height=14cm,width=12cm,viewport=40 120 570 725,clip]{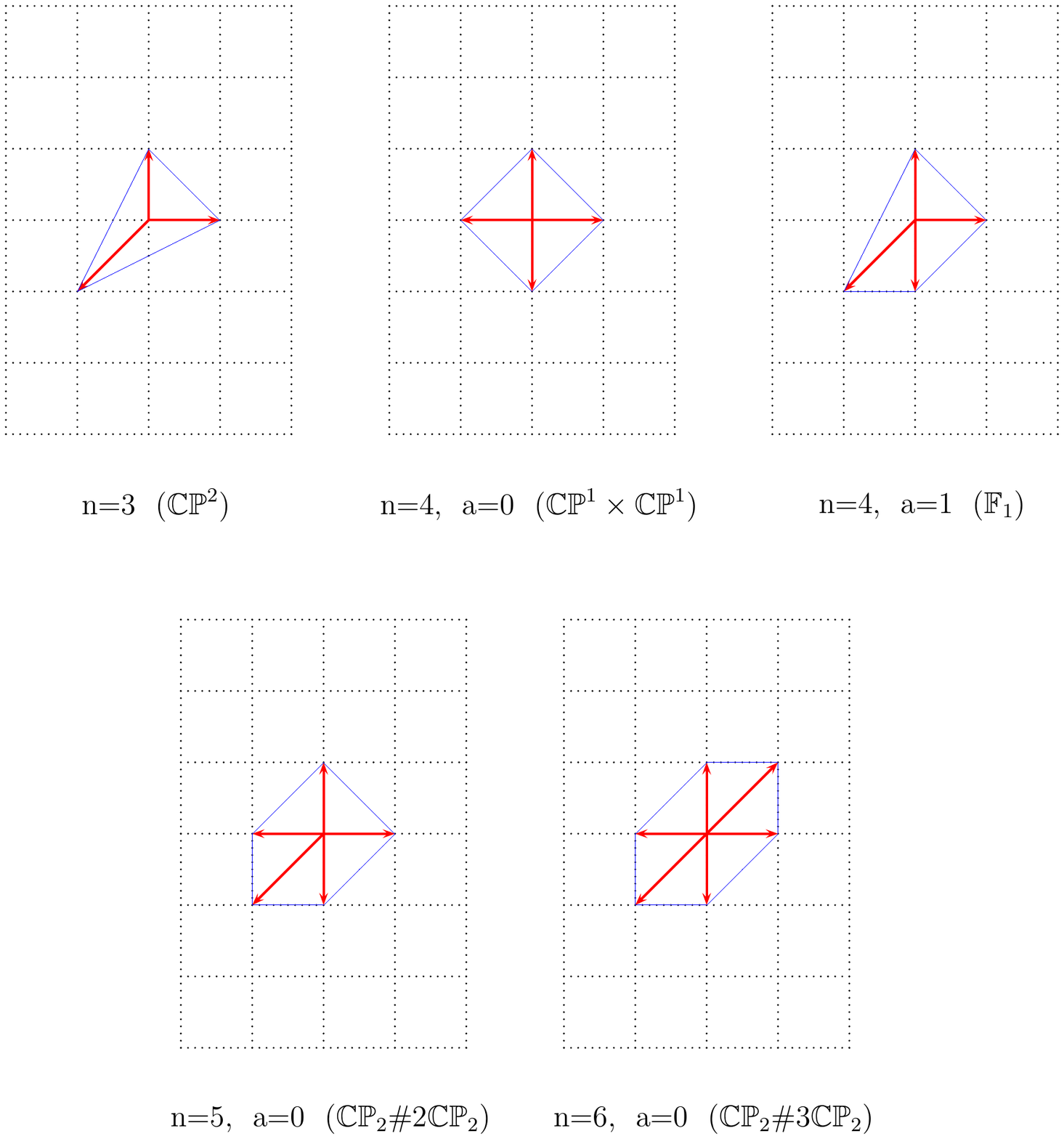}
\caption{Two-dimensional fans $\Sigma$ and fundamental 
generators corresponding to smooth toric Fano surfaces.}
\label{2dfano}
\end{figure}

All admissible circular graphs with number of vertices $n\leq6$ are shown in fig.~\ref{circular}. Performing the exercise described above for these graphs  we obtain the following result:
\begin{itemize}
\item $n=3$: $G(\Sigma)=\{e_1,e_2,-e_1-e_2 \}$. This is the two-dimensional projective space $\mathbb{CP}^2$.
\item $n=4$: $G(\Sigma)=\{e_1,e_2,-e_1-a e_2,-e_2 \}$. For $a=0$ this is 
$\mathbb{CP}^1\times\mathbb{CP}^1$ (also known as the Hirzebruch surface $\mathbb{F}_0$); for $a=1$ it is the Hirzebruch surface $\mathbb{F}_1$ ($\mathbb{CP}^2$ blown up at one point). 
\item $n=5$: $G(\Sigma)=\{e_1,e_2,-e_1-a e_2,-e_1-(a+1) e_2,-e_2 \}$. For 
$a=0$ this is the del Pezzo surface $\mathbb{CP}^2\#2\mathbb{CP}^2$ ($\mathbb{CP}^2$ blown up at two points).
\item $n=6$ (1st): $G(\Sigma)=\{e_1,e_1+e_2,e_2,-e_1-(a+1) e_2,-e_1-(a+2) e_2,-e_2 \}$. For $a=-1$ this is the del Pezzo surface $\mathbb{CP}^2\#3\mathbb{CP}^2$ ($\mathbb{CP}^2$ blown up at three points).
\item $n=6$ (2nd): $G(\Sigma)=\{e_1,e_2,-e_1-a e_2,-2e_1-(2a+1)e_2,-e_1-(a+1) e_2,-e_2 \}$
\item $n=6$ (3rd): $G(\Sigma)=\{e_1,e_2,-e_1-a e_2,-e_1-(a+1)e_2,-e_1-(a+2) e_2,-e_2 \}$
\end{itemize}
where in all the cases above $\{e_1,e_2 \}$ is a $\mathbb{Z}$-basis of $N$ 
and  $a\in\mathbb{Z}$. The corresponding fans $\Sigma$ and fundamental generators $G(\Sigma)$ are shown in fig.~\ref{2d}. Of these, the subset corresponding to the five two-dimensional smooth Fano varieties are depicted in fig.~\ref{2dfano}.

Let us finally mention for completeness that, as shown in \cite{ota}, all two-dimensional SCTV can be obtained by either $\mathbb{CP}^2$ or the Hirzebruch surface $\mathbb{F}_a$, $a=0,2,3,\dots$, by a finite number of blow-ups.

\subsection*{Three-dimensional classification}

Let $d=3$ and $N\cong\mathbb{Z}^3$. Let $S$ be a two-sphere centered at the 
origin of $N$ and consider a triangulation of $S$ by spherical triangles. First we will need a number of definitions:

\begin{itemize}
\item An {\it $N$-weighting} of the triangulation is an assignment of a primitive element $v\in N$ to each spherical vertex. 
\item A {\it double $\mathbb{Z}$-weighting} of the triangulation is an 
assignment of a pair of integers to each spherical edge, such that 
one integer is on the side of one vertex and the other integer is on the 
side of the other vertex. 
Fig.~\ref{sphericaltriangle} depicts two spherical triangles $v_1v_2v$ and $v_1v_2v$, together with the primitive elements of $\nr$ assigned to the vertices. The $\mathbb{Z}$-weights $a$, $b$ of the edge $v_1v_2$ are also depicted,  with $a$ on the side of $v_1$ and $b$ on the side of $v_2$.
\begin{figure}
\centering
\includegraphics[height=4.3cm,width=4.3cm,viewport=45 165 495 615,clip]{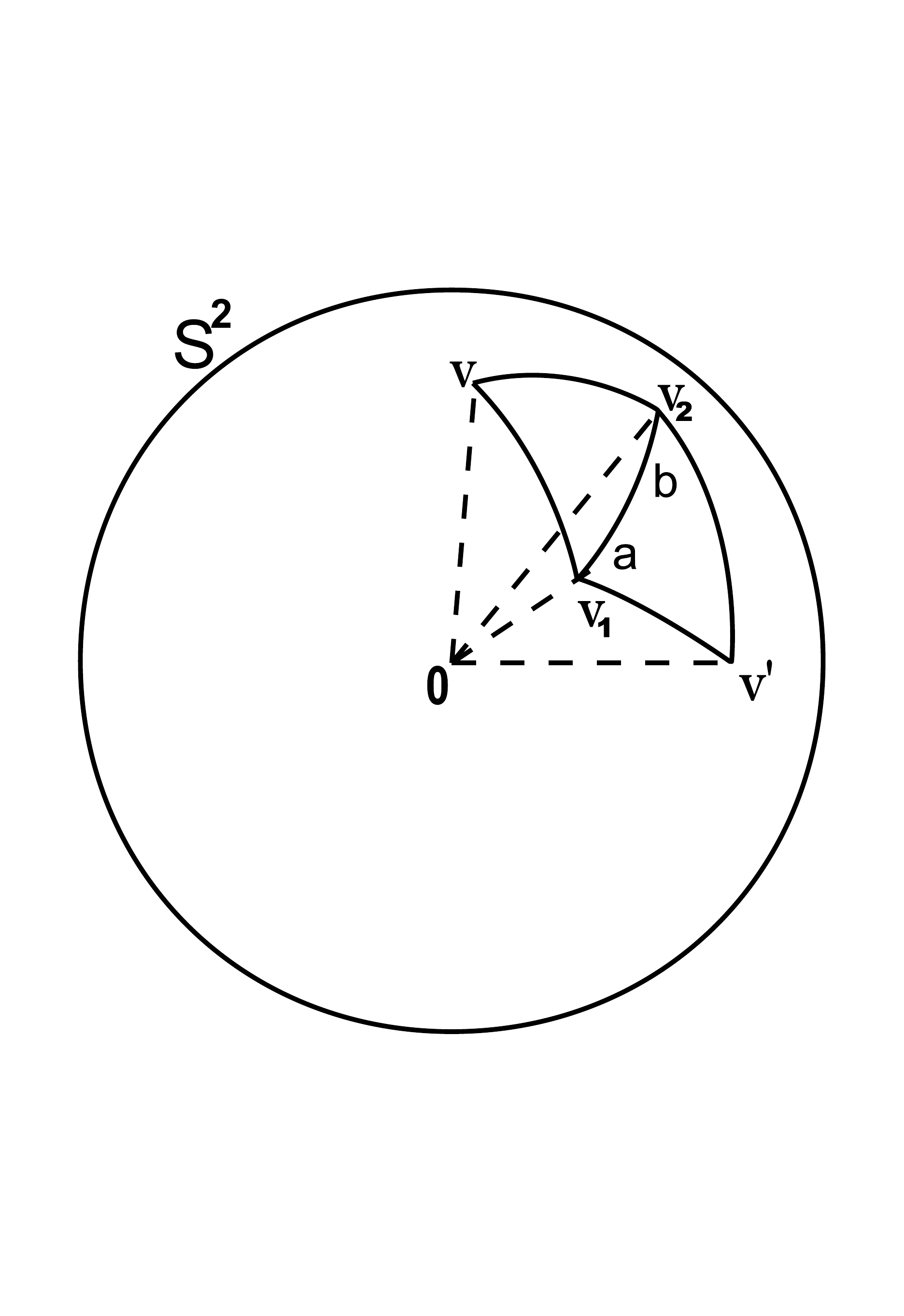}
\caption{Two adjacent  spherical triangles $v_1v_2v$ and $v_1v_2v'$.  The $\mathbb{Z}$-weights $a$, $b$  of the common spherical edge $v_1v_2$ are also depicted.
}
\label{sphericaltriangle}
\end{figure}
\item Let $\{v_1,\dots,v_r\}$ be the vertices adjacent to an $r$-valent vertex $v$, and let $a_i$ be the $\mathbb{Z}$-weight assigned to the spherical edge $vv_i$ which is on the side of $v_i$. The {\it weighted link  of $v$} is the spherical polygon $v_1v_2\dots v_rv_1$ together with the weights $a_i$, $i=1,\dots,r$. This is depicted in fig.~\ref{weightedlink}. 
\begin{figure}
\centering
\includegraphics[height=4.3cm,width=4.7cm,viewport=50 170 495 600,clip]{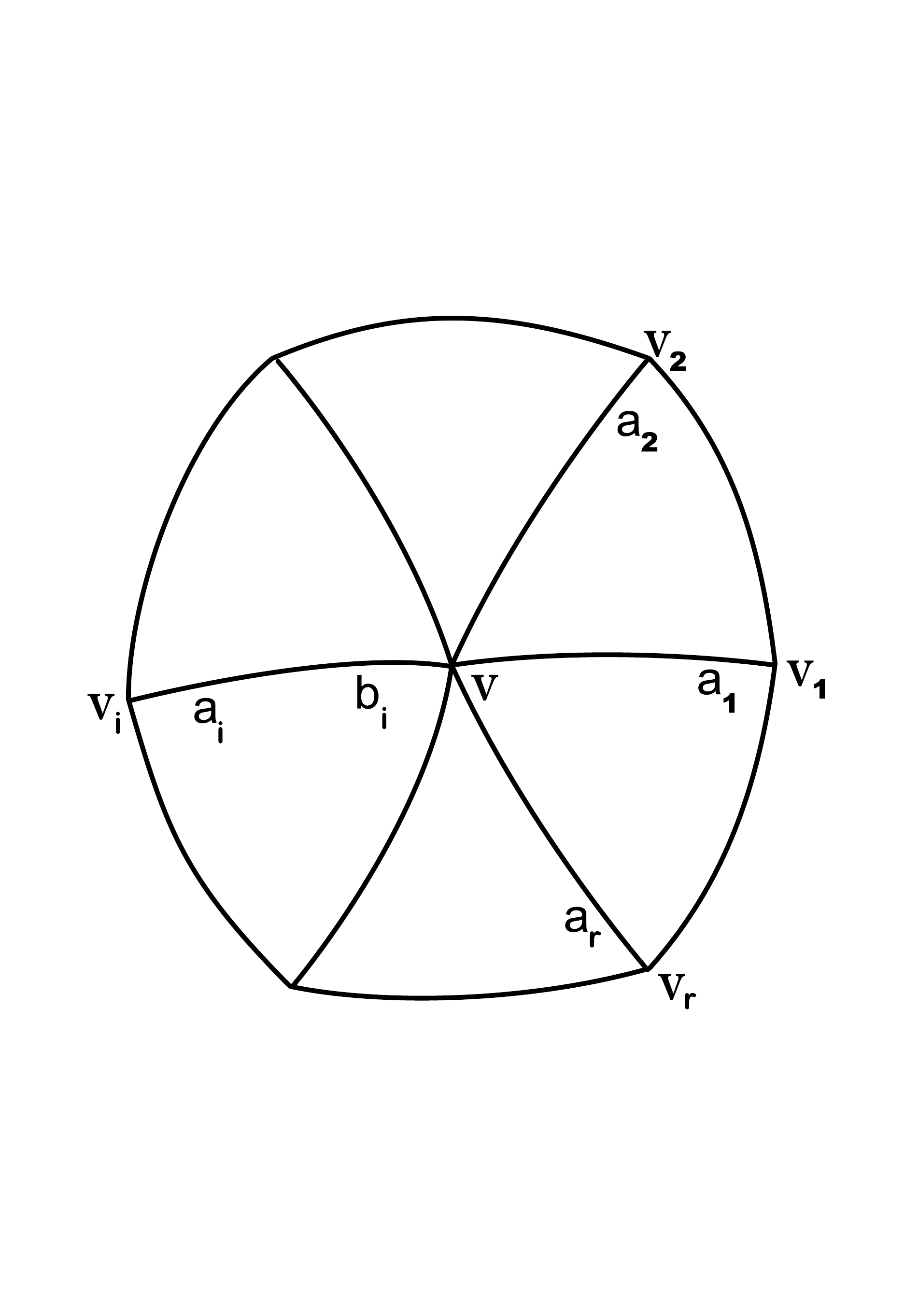}
\caption{The weighted spherical link of an $r$-valent vertex $v$.}
\label{weightedlink}
\end{figure}
\item By intersecting a fan $\Sigma$ with the sphere $S$, one obtains a triangulation of $S$ together with a canonical $N$-weighting: to each vertex of the triangulation we assign the fundamental generator of the corresponding 
one-dimensional cone in $\Sigma$. If the fundamental 
generator is $v$, we will call that vertex {\it the vertex $v$}. (Note however, that the vector $v$ itself need not be on $S$). An {\it admissible $N$-weighting} is an  $N$-weighting which is obtained in the way just described.

\end{itemize}

An admissible $N$-weighting gives rise to a double $\mathbb{Z}$-weighting. Indeed, consider the spherical edge $v_1v_2$ corresponding to the intersection 
of $S$ with the two dimensional cone $\tau$ generated by $v_1$, $v_2$ in $\Sigma$. By the same reasoning that led to eq.~(\ref{central}), there exist exactly two maximal cones $\sigma$ and $\sigma'$, generated by $\{v_1,v_2,v\}$ and $\{v_1,v_2,v'\}$ respectively, such that $\tau$ is the common facet 
of $\sigma$ and $\sigma'$. Moreover there exist $a$, $b\in\mathbb{Z}$ such that:
\eq{ \label{central3}
v+v'+a v_1+b v_2=0~.
}
We then assign the pair $(a,b)$ to the edge $v_1v_2$, with $a$ on the 
side of $v_1$ and $b$ on the side of $v_2$ (see fig.~\ref{sphericaltriangle}). 
We obtain the $\mathbb{Z}$-weighting by repeating this procedure for each 
spherical edge. More generally, consider the link of a vertex $v$ with the 
adjacent vertices $\{v_1,\dots,v_r\}$, as in fig.~\ref{weightedlink}. 
Applying eq.~(\ref{central}) to this case, there exist pairs of integers $(a_i,b_i)$, such that 
\eq{ \label{central3i}
v_{i-1}+v_{i-1}+a_i v_i+b_i v=0~,
}
for $i=1,\dots,r$.

\begin{itemize}

\item A double $\mathbb{Z}$-weighting is called  {\it admissible} if (i) eqs.~(\ref{central3i}) are compatible with each other as $v$ runs through 
all vertices in the triangulation and (ii) the 
weighted link of each vertex $v$ is an admissible weighted circular graph.

\end{itemize}
In \cite{ota} the classification of three-dimensional SCTV 
was shown to reduce to the classification of triangulations of $S$ with admissible weightings. More precisely, it was shown that:
\begin{center}
\framebox{\parbox[t]{14.5cm}{\it
There exist canonical isomorphisms  between: (i) three-dimensional SCTV (ii) admissible double $\mathbb{Z}$-weightings of $S$ (iii) admissible $N$-weightings of $S$.  
}}
\end{center}

\subsubsection*{Three-dimensional smooth, compact toric $\mathbb{CP}^1$  bundles}

For triangulations of $S$ with up to eight vertices, \cite{ota} provided 
the complete list of corresponding admissible weightings. We will mainly be interested in one particular class thereof: it is 
`case (2)' of the triangulation denoted by [$n-1$] in \cite{ota} (see Theorem 9.6 and Lemma 9.7 therein), where $n$ is the number of vertices of the triangluation. This is merely for convenience since, as we will now see, the corresponding three-dimensional SCTV can be concisely described without any restriction on the number of vertices $n$ of the triangulation. We stress that our 
methods are also applicable to any of the other classes of SCTV.

Before we describe these toric $\mathbb{CP}^1$  bundles in more detail, 
let us say a few words about the remaining SCTV in the classification: 
`Case (1)' of the triangulation denoted by [$n-1$] in \cite{ota} corresponds to 
$\mathbb{CP}^2$ bundles over $\mathbb{CP}^1$ whose `twisting' is 
characterized by two integers -- we will describe these explicitly using 
the symplectic quotient description in section \ref{symplectic}. In addition, 
\cite{ota} derived all admissible weightings for triangulations with number of vertices $n\leq 8$ as well as for the 
triangulation with $n=12$ induced by the eicosahedron.

Let us now proceed to the description of our chosen class of SCTV: 
Their corresponding fans are generated by $\{v_1,\dots,v_n\}$ which 
may be chosen such that
\eq{
v_1+v_2=0~,
}
while the images of the remaining generators  $\{v_3,\dots,v_n\}$ in $N/\mathbb{Z}\otimes v_1$ 
can be identified with the set of generators of a two-dimensional SCTV. 
As it turns out, these  SCTV are $\mathbb{CP}^1$ bundles over 
two-dimensional SCTV, with the fiber generated by $\{v_1,v_2\}$. This 
will become more explicit when we give their symplectic 
quotient description in section \ref{symplectic}.

\begin{figure}
\centering
\includegraphics[height=4.3cm,width=4.7cm,viewport=50 170 495 600,clip]{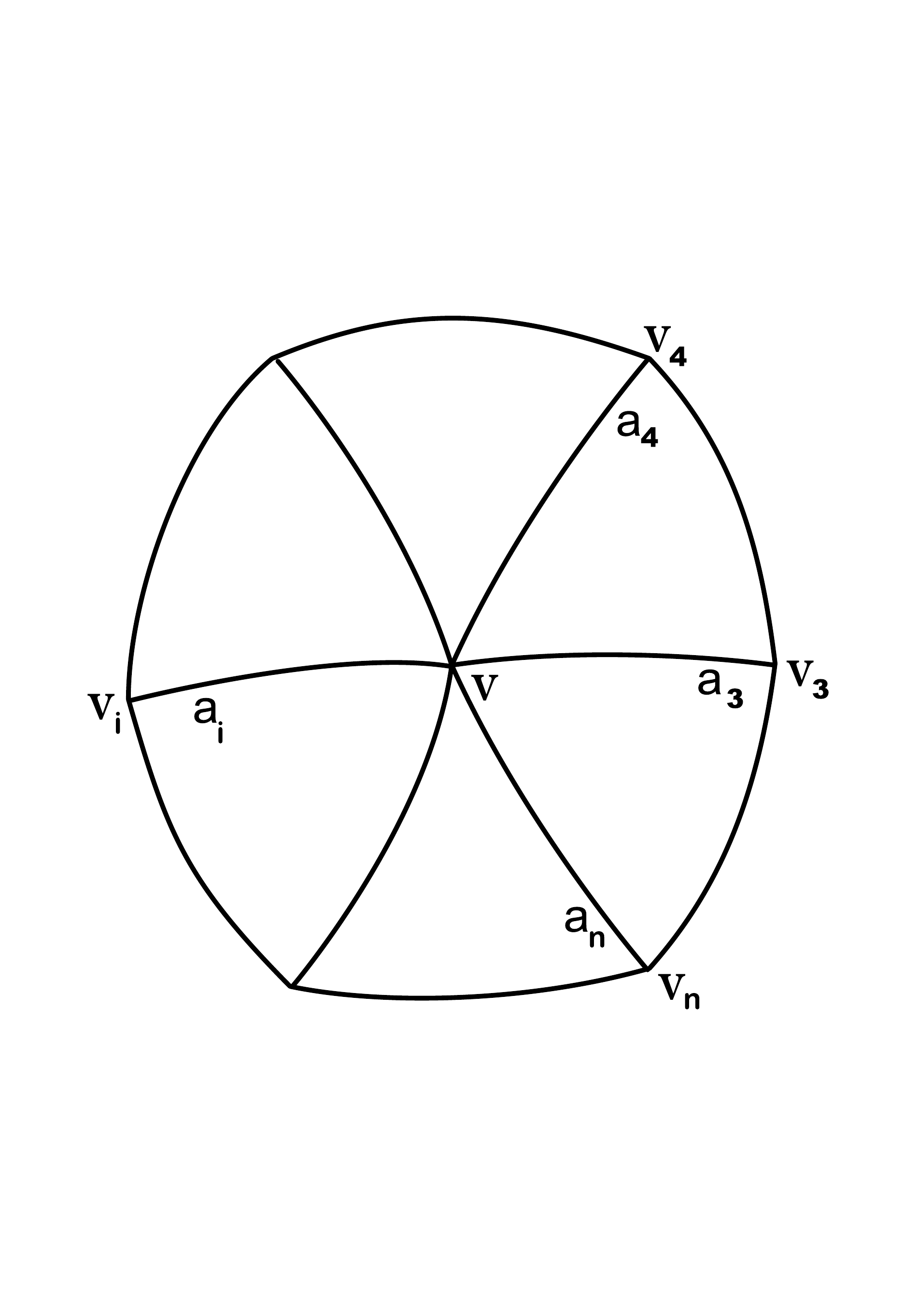}
\caption{The weighted spherical link of the vertex $v=v_1$ or $v=v_2$ 
of a toric $\mathbb{CP}^1$ bundle.}
\label{weightedlinkp1}
\end{figure}

\begin{figure}
\centering
\includegraphics[height=6.2cm,width=5cm,viewport=25 150 565 695,clip]{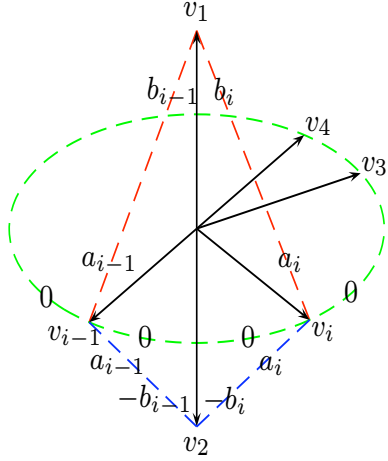}
\caption{Three-dimensional double $\mathbb{Z}$-weighting for
smooth, compact toric $\mathbb{CP}^1$ bundles.}
\label{3dcircular}
\end{figure}

Given the above information, it is straightforward to construct 
the corrseponding double $\mathbb{Z}$-weighting for these $\mathbb{CP}^1$ bundles: The weighted links of $v_1$ and $v_2$ are admissible 
weighted circular graphs of some (the same for  both $v_1$ and $v_2$) 
two-dimensional SCTV (\ie{} they are of the form of   
fig.~\ref{weightedlinkp1} for $v\rightarrow v_1$ and  $v\rightarrow v_2$ respectively); the weighted link of $v_i$, where $i=3,\dots,n$, is 
of the form of  the $n=4$ graph in fig.~\ref{circular} for $a\rightarrow b_i$. All this is depicted in fig.~\ref{3dcircular}.

\subsection{Symplectic quotient description}\label{symplectic}

In the previous section we focused on the description of SCTV in terms 
of fans, and reviewed their classification in two and three dimensions. 
In this section we will introduce the equivalent symplectic-quotient description 
of toric varieties, which turns out to be more useful for the construction 
of $G$-structures (which is 
the subject of section \ref{vacua} and the ultimate 
goal of this paper), and explain how it is related to the description of section \ref{toricreview}.

Let $\{z^i, ~i=1,\dots n\}$, be holomorphic coordinates of $\mathbb{C}^n$. 
Moreover, consider a $U(1)^s$ action on $\mathbb{C}^n$ given by:
\eq{\label{u1action}
z^i\longrightarrow e^{i\varphi_aQ^a_i}z^i~,
}
where $\{Q^a_i,~ a=1,\dots s\}$ are the $U(1)$ charges. 
A { toric variety} of real dimension $2d$  can be  defined 
as the quotient
\eq{\label{kmoddef}
\mathcal{M}_{2d}=\{
z^i\in\mathbb{C}^n | \sum_{i=1}^nQ^a_i|z^i|^2=\xi^a
\}/U(1)^s~,
}
where $d=n-s$ and the $U(1)^s$ acts as in (\ref{u1action}). 
It can be shown that the Fayet-Iliopoulos parameters $\xi^a$  
are in fact the K\"{a}hler moduli of  $\mathcal{M}_{2d}$ (see e.g. \cite{denef}).

Consider the holomorphic vector fields 
\eq{\label{holvec}
V^a:=\sum_{i}Q^a_iz^i\partial_{z_i}
}
generating the $U(1)^s$ action. 
A form $\Phi$ on $\mathbb{C}^n$ descends to 
a well defined-form on  $\mathcal{M}_{2d}$, henceforth denoted by $\Phi|$,  if it is {\it vertical}~\!:
\eq{\label{vertical}
\iota_{V^a}\Phi=\iota_{\bar{V}^a}\Phi=0~,
}
for $a=1,\dots, s$, 
and {\it invariant}~\!:
\eq{\label{invariant}
\mathcal{L}_{\mathrm{Im}V^a}\Phi=0~,
}
where $\mathcal{L}_{V}$ is the Lie derivative with respect to the 
vector $V$.  Note that the condition of invariance (\ref{invariant}) is 
by definition equivalent to the condition of {\it gauge-invariance} 
under $U(1)^s$ transformations.

The above conditions can be rephrased in the language of a {\it 
symplectic quotient}. Let us introduce the {\it moment maps}:
\eq{
\mu^a:= Q_i^a|z^i|^2
~.}
The toric variety is given by the quotient:
\eq{\label{quotient}
\mathcal{M}_{2d}=\mu^{-1}(\xi)/U(1)^s
~,}
where the $U(1)^s$ action was defined in (\ref{u1action}). 

We can 
understand the conditions (\ref{vertical},\ref{invariant}) 
in the language of the symplectic quotient as follows: 
A form $\Phi$ on $\mathbb{C}^n$ restricts to a well-defined
 form  on 
$\mu^{-1}(\xi)$, provided it does not have any components along 
$d\mu^a$. This is equivalent to the condition that $\Phi$ be vertical 
with respect to vector $\mathrm{Re}V^a$ 
(which is the dual of $d\mu^a$): $\iota_{\mathrm{Re}V^a}\Phi=0$. 
We can then mod out by the action of $U(1)^s$ 
as in (\ref{quotient}) provided $\Phi$ is gauge-invariant, $\mathcal{L}_{\mathrm{Im}V^a}\Phi=0$, and  
has no components along the orbits of $\mathrm{Im}V^a$, 
$\iota_{\mathrm{Im}V^a}\Phi=0$.  
Furthermore, if $\Phi$ descends to 
a well defined-form  $\Phi|$ on  $\mathcal{M}_{2d}$, 
$\Phi|$ is invariant under $\mathrm{Re}V^a$. Indeed,  
since  $d(\Phi|)$  has no components along $d\mu^a$ we have
\eq{\label{rest}
0=\iota_{\mathrm{Re}V^a}d(\Phi|)
=\left(\iota_{\mathrm{Re}V^a}d+d\iota_{\mathrm{Re}V^a}\right)\left(\Phi|\right)
=\mathcal{L}_{\mathrm{Re}V^a}\left(\Phi|\right)
~,}
where in the second equality we used that $\Phi|$ is vertical.

It is useful to introduce the 
$(1,0)$-forms $\eta^a$ dual to the holomorphic vectors 
defined in (\ref{holvec}):
\eq{\label{holf}
\eta^a:=Q^a_i\bar{z}_idz^i
~.}
Let us also introduce the real symmetric matrix $g$
\eq{\label{gdef}
g_{ab}:=\iota_{V^a}\eta^b=Q^a_iQ^b_i|z^i|^2
}
and its inverse $\tilde{g}$
\eq{
g_{ae}\tilde{g}_{eb}=\delta_{ab}
~.}
We can now define a {\it vertical projector} $P$. 
For a $(k,0)$ form $\Phi$: %
\eq{
P(\Phi)=\Phi
+\sum_{n=1}^{k} 
\frac{(-1)^{ \frac{1}{2}n(n+1) }  }{n!} 
\tilde{g}_{a_{1}a_1'}\dots \tilde{g}_{a_n a_n'}
\eta^{a_1}\wedge\dots\wedge\eta^{a_ n}\iota_{V^{a_1'}}\dots\iota_{V^{a_n'}}\Phi
~,}
and similarly for a general $(k,l)$ form. 
In other words, by definition $P$ projects out all components along $\eta$, $\bar{\eta}$.

If $\Phi$ descends to 
a well-defined form $\Phi|$ on  $\mathcal{M}_{2d}$, then 
the following useful identity facilitates the computation 
of the exterior differential:
\eq{\label{arm}
d(\Phi|)=P(d\Phi)|
~.}
This can be seen by noting that $(d\Phi)|$ potentially 
differs from $d(\Phi|)$ by additional components 
along $d\mu^a=2\mathrm{Re}\eta^a$.

As in \cite{gt} let us define the vertical component of $dz^i$:
\eq{
\mathcal{D}z^i:=P(dz^i)=dz^i- \tilde{g}_{ab}\eta^aQ^b_iz^i~.
}
We note that
\eq{\label{calderiv}
P(d\mathcal{D}z^i)=
h_{ik}z^i\mathcal{D}z^k\wedge\mathcal{D}\bar{z}_k
~,}
(no summation over $i$) where
\eq{\label{hdef}
h_{ij}:=Q^a_iQ^b_j\tilde{g}_{ab}
~.}
Using the above we can also define $\mathcal{D}z^i$ as follows:
\eq{
\mathcal{D}z^i=P_{ij}dz^j
~,}
where $P_{ij}$ is the component form of the 
vertical projector acting on (1,0)-forms:
\eq{\label{projector}
P_{ij}=(P_{j~\!\!i})^*
=\delta_{ij}-h_{ij}z^i\bar{z}_j
~.}
The projector property 
\eq{
P_{ik}P_{kl}=P_{il}
~,}
can be shown using the identity
\eq{\label{us1}
h_{ik}h_{kl}|z_k|^2=h_{il}
~,
}
which is a consequence of definitions (\ref{gdef},\ref{hdef}). 

The K\"{a}hler form of $\mathbb{C}^n$ 
\eq{\label{jcdef}
J_{\mathbb{C}}:=\frac{i}{2} dz^i\wedge d\bar{z}_i
}
projects to its vertical component:
\eq{\spl{\label{jdef}
\widetilde{J}&:=P(J_{\mathbb{C}})\\
&~=\frac{i}{2}dz^i\wedge d\bar{z}_i
+\frac{i}{2}\tilde{g}_{ab}\eta^a\wedge\bar{\eta}^b\\
&~=\frac{i}{2}\mathcal{D}z^i\wedge \mathcal{D}\bar{z}_i
~,}}
which descends to a well-defined form $\widetilde{J}|$ on $\mathcal{M}_{2d}$. 
Taking  (\ref{arm}) into account it is easy to see that $\widetilde{J}|$ is closed.

\subsection*{Relation to the previous description}

Given the set of fundamental generators $G(\Sigma)$ of a fan $\Sigma$ 
corresponding to a variety $V_\Sigma$ it is straightforward to pass to the symplectic quotient description: Since the fundamental generators 
$\{v_1,\dots,v_n\}$ are vectors in $\nr\cong\mathbb{R}^d$, there exist 
$n-d$ linear relations among them; they are exactly provided by the 
$U(1)$ charges $Q^a$,
\eq{\label{connection}
\sum_{i=1}^nQ_i^{a} v_i=0~,
}
for $a=1,\dots,s=n-d$. 

Eq.~(\ref{connection}) provides the connection between the two descriptions:  given the set of fundamental generators $G(\Sigma)$, we can solve it in order 
to determine  the $U(1)$ charges $Q^a$.~\!\footnote{The basis of $U(1)$ charges is not uniquely 
determined. The ambiguity may be partially eliminated by taking the two-cycles 
$\mathcal{C}^a$ corresponding to $Q^a$ to be generators of the Mori cone, 
see e.g. \cite{denef}. As an aside, recall that for the variety $V_\Sigma$ to be CY (and thus have vanishing first Chern class), the sum  $\sum_iQ_i^{a}$ must vanish for all charges. } Performing this exercise for the 
sets of generators of two-dimensional fans computed in section \ref{classification}, corresponding to the admissible weighted circular graphs of fig.~\ref{circular}, we obtain the following result:
\begin{itemize}
\item $n=3$: $Q^1=(1,1,1)$.
\item $n=4$: $Q^1=(0,1,0,1)$, $Q^2=(1,a,1,0)$. 
\item $n=5$: $Q^1=(1,a,1,0,0)$, $Q^2=(0,1,0,0,1)$, $Q^3=(1,a+1,0,1,0)$. 
\item $n=6$ (1st): $Q^1=(-1,1,-1,0,0,0)$, $Q^2=(1,0,a,1,0,0)$, $Q^3=(0,0,1,0,0,1)$, $Q^4=(1,0,a+1,0,1,0)$. 
\item $n=6$ (2nd): $Q^1=(1,a,1,0,0,0)$, $Q^2=(2,2a+1,0,1,0,0)$, $Q^3=(1,a+1,0,0,1,0)$, $Q^4=(0,1,0,0,0,1)$. 
\item $n=6$ (3rd): $Q^1=(1,a,1,0,0,0)$, $Q^2=(1,a+1,0,1,0,0)$, $Q^3=(1,a+2,0,0,1,0)$, $Q^4=(0,1,0,0,0,1)$. 
\end{itemize}
The construction of the $U(1)$ charges corresponding to the toric $\mathbb{CP}^1$ bundles described at the end of section \ref{classification} is also straightforward to derive: Consider the double $\mathbb{Z}$-weighting 
of fig.~\ref{3dcircular}. Let the weighted link of $v_1$ (also $v_2$) 
correspond to  a two dimensional SCTV  with associated $U(1)$ charges 
given by $q^a$, $a=1,\dots,s-1$. By taking eqs.~(\ref{central3i}) into account, it is not difficult to see that eqs.~(\ref{connection}) 
are solved by the following charges:
\eq{\spl{
Q^a&=(q^a,c_a,0)~~;~~~~~a=1,\dots,s-1~,\\
Q^a&=(0,\dots,0,1,1)~~;~~~~~a=s~,
}}
where $c_a\in\mathbb{Z}$, $a=1,\dots,s-1$, are integers specifying the 
`twisting' of the $\mathbb{CP}^1$ bundle.

To illustrate this with an example, let us take  the toric $\mathbb{CP}^1$ bundle 
over $\mathbb{CP}^2$. In this case the weighted link of $v_1$ is the 
$n=3$ graph of fig.~\ref{circular}, with corresponding $U(1)$ charge $q=(1,1,1)$ (as shown in section \ref{classification}). Hence the $U(1)$ charges 
for the $\mathbb{CP}^1$ over $\mathbb{CP}^2$ bundle read $Q^1=(1,1,1,c,0)$, $Q^2=(0,0,0,1,1)$, with  $c\in\mathbb{Z}$.

For completeness let us also list the $U(1)$ charges of the class 
of $\mathbb{CP}^2$ bundles over $\mathbb{CP}^1$ mentioned in section \ref{classification} (`case (1)' of the triangulation denoted by [$n-1$] in \cite{ota}). These are given by:
\eq{
Q^1=(1,1,a,b,0)~~;~~~~~Q^2=(0,0,1,1,1)~,
}
where $a$, $b$ are integers specifying the 
`twisting' of the $\mathbb{CP}^2$ bundle. We also note that these are special cases of the more general $d$-dimensional toric varieties with $n=d+2$ generators 
constructed in \cite{klein}.

\section{Toric vacua}\label{vacua}

After a brief review of $SU(3)$ structures, 
we will now outline a procedure which allows one 
to construct $SU(3)$ structures on 
three-dimensional SCTV $\mathcal{M}_6=V_\Sigma$. 
Our construction requires the existence of a (1,0)-form 
$K$ on the parent space $\mathbb{C}^n$ obeying certain conditions   
listed in section \ref{sec:4.2}. This procedure is then illustrated in 
section \ref{examples} with a couple of explicit examples: the first one 
is $\mathbb{CP}^3$ and has appeared before in the literature; 
the second one is new and is an infinite class of toric 
$\mathbb{CP}^1$ bundles.

\subsection{Review of $SU(3)$ structures}\label{su3a}

Let $\cm_6$ be a six-dimensional, smooth manifold. Let us assume that 
$\cm_6$ is equipped with a (Riemannian) metric $g$ and an almost-complex structure $\mathcal{I}$ with associated 
two-form $J$. The triplet $(\cm_6,g,J)$ defines a $U(3)$ structure on $\cm_6$. An {\it $SU(3)$ structure}  
on $\cm_6$ is obtained if, in addition,  a non-degenerate complex (3,0) form $\Omega$ exists on $\cm_6$ .

Equivalently, we can define an  $SU(3)$ structure on $\cm_6$ as the triplet $(\cm_6,J,\Omega)$, where 
$J$ is a real two-form and $\Omega$ is a complex decomposable three-form such that:
\boxedeq{\label{su3}
\spl{
\Omega\wedge J&=0 \, , \\
\Omega\wedge\Omega^*&=\frac{4i}{3}J^3\neq 0
~.
}}
In this second formulation, the metric is determined from $(J,\Omega)$ as follows \cite{hitc}:  
Let us define
\eq{
\tilde{\mathcal{I}}_k{}^l = \varepsilon^{lm_1\dots m_5} (\Re \Omega)_{km_1m_2} (\Re \Omega)_{m_3m_4m_5} \, ,
}
where $\varepsilon^{m_1\dots m_6}=\pm1$ is the totally antisymmetric symbol in six dimensions. The condition that $\Omega$ be complex decomposable can only be fulfilled if $\tilde{\mathcal{I}}^2$ is strictly negative. 
In this case an almost complex structure can be defined  by
\eq{\label{hitcncl}
\mathcal{I} = \frac{\tilde{\mathcal{I}}}{\sqrt{-\text{tr}\, \frac{1}{6}\,\tilde{\mathcal{I}}^2}} \, ,
}
which is properly  normalized so that $\mathcal{I}^2 = -\bbone$,  
and the complex-decomposability of  $\Omega$ 
amounts to requiring that $\mathrm{Im}\Omega$ be related to  $\mathrm{Re}\Omega$
via 
\eq{
\mathrm{Im}\Omega=\pm\frac{1}{3}\mathcal{I}_k{}^ldx^k\wedge \iota_l\mathrm{Re}\Omega~.
}

The metric can be constructed from $\mathcal{I}$ and $J$ via
\eq{
\label{su3metric}
g_{mn}=\mathcal{I}_m{}^lJ_{ln}~.
}
Positivity has to be additionally imposed, if we want $g$ to be Riemannian.

The {\it torsion classes} of $(\cm_6,J,\Omega)$ are defined through the decomposition 
of the exterior differentials of $J$, $\Omega$ into $SU(3)$-modules. 
Intuitively, the  torsion classes parameterize the 
failure of the manifold to be of special holonomy, which can also
be thought of as the failure of $J$, $\Omega$ to be closed. 
More specifically we have  \cite{cs,zoup}:
\eq{\label{torsionclass}
\spl{
d J&=\frac{3}{2}\Im(\mathcal{W}_1\Omega^*)+\mathcal{W}_4\wedge J+\mathcal{W}_3 \, , \\
d \Omega&= \mathcal{W}_1 J\wedge J+\mathcal{W}_2 \wedge J+\mathcal{W}_5^*\wedge \Omega ~,
}}
where $\mathcal{W}_1$ is a function, $\mathcal{W}_2$ is a primitive (1,1)-form, $\mathcal{W}_3$ is a real
primitive $(1,2)+(2,1)$-form, $\mathcal{W}_4$ is a real one-form and $\mathcal{W}_5$ a complex (1,0)-form.

\subsection{$SU(3)$ structures on SCTV}\label{sec:4.2}

Let us start by noticing that the basis of (1,0)-forms (with respect to 
the induced complex structure from $\mathbb{C}^n$) $\mathcal{D}z^i$, $i=1,\dots,n$,  defined in section \ref{symplectic} is redundant as there cannot exist more than  $d$ independent (1,0)-forms on $\mathcal{M}_{2d}$. Indeed, this can be seen directly from the fact that the  $\mathcal{D}z^i$'s obey the following $d-n$ conditions:
\eq{\label{35}
\sum_{i=1}^nQ^a_i\bar{z}_i\mathcal{D}z^i~;~~~a=1,\dots,d-n~.
}
These equations follow from $P(\eta^a)=0$, by taking the definition of $\eta^a$,  eqn.~(\ref{holf}), into account. Note that this precisely parallels  eqn.~(\ref{connection}) with the well-defined (\ie{} vertical and  gauge-invariant) one-forms $\bar{z}_i\mathcal{D}z^i$ on $\mathcal{M}_{2d}$ replacing the one-dimensional cone generators $v_i$. Moreover 
the consistency condition 
\eq{
dP(\eta^a)=0~,
}
can easily be seen to follow from eqns.~(\ref{arm},\ref{calderiv}) upon taking 
eqns.~(\ref{gdef},\ref{hdef}) into account.

The toric manifold $\mathcal{M}_{2d}$ is naturally equipped with a {\it local} 
$SU(d)$ structure. This follows from the fact that it is K\"{a}hler and admits a global $U(d)$ structure. To construct the local $SU(d)$ structure explicitly 
one has to supplement the induced K\"{a}hler form $\widetilde{J}\vert$ on $\mathcal{M}_{2d}$ eq.~(\ref{jdef}), with a complex $d$-form $\widetilde{\Omega}\vert$, where 
\eq{\label{omans}
\widetilde{\Omega}:= \left(\mathrm{det}g_{ab}\right)^{-1/2}\prod_{a=1}^{s}\iota_{V^{a}}
\Omega_{\mathbb{C}}~,
}
and $\Omega_{\mathbb{C}}$ is the holomorphic top form of $\mathbb{C}^n$. The reason for the above normalization will become clear momentarily.

The pair $(\widetilde{J}, \widetilde{\Omega})$ satisfies 
the orthogonality condition
\eq{\label{orth1}
\widetilde{J}\wedge \widetilde{\Omega}=0
~.}
This follows from 
\eq{\label{jorthb}
J_{\mathbb{C}}\wedge{\Omega}_{\mathbb{C}}=0
~,}
upon contracting with $\prod_{a=1}^{s}\iota_{V^{a}}$, projecting with $P$ onto $\cm_{2d}$, and noticing that 
\eq{
P(\iota_{V^a}J_{\mathbb{C}})=P(\bar{\eta}^a)=0~.
}
Moreover the normalization condition
\eq{\label{gennorm}
i^d\widetilde{J}^d\propto  
\widetilde{\Omega}\wedge \widetilde{\Omega}^*
~,}
follows from 
\eq{\label{dgennorm}
i^n J_{\mathbb{C}}^n\propto {\Omega}_{\mathbb{C}}\wedge{\Omega}_{\mathbb{C}}^*
~,}
up to real proportionality constants. 
Indeed, upon contracting with $\prod_{a,b=1}^{s}\iota_{V^{a}}\iota_{\bar{V}^{b}}$, projecting with $P$, and taking (\ref{gdef},\ref{jdef}) into account, we find that the contraction of 
the left-hand side above 
is proportional to $\mathrm{det}g_{ab}~\! i^d\widetilde{J}^d$. Also, by taking 
(\ref{35}) into account the contraction of right hand side gives  
$\mathrm{det}g_{ab}~\!\widetilde{\Omega}\wedge \widetilde{\Omega}^*$, provided that $\widetilde{\Omega}$ is normalized as in \eqref{omans}

The complex $d$-form $\widetilde{\Omega}\vert$  is regular on  $\cm_{2d}$ (it has no poles) and moreover is vertical.  It therefore follows from the above construction that the pair $(\widetilde{J}\vert, \widetilde{\Omega}\vert)$ defines a local $SU(d)$ 
structure on  $\cm_{2d}$, with associated metric induced from 
the canonical metric on $\mathbb{C}^n$. However, $(\widetilde{J}\vert, \widetilde{\Omega}\vert)$ fails to be a global $SU(d)$ 
structure on  $\cm_{2d}$ because $\widetilde{\Omega}\vert$  is not  gauge-invariant (it has non-zero $Q^a$ charge).

Let us now specialize to $d=3$. 
One can construct a natural ansatz for a global $SU(3)$ 
structure on  $\cm_{6}$ by appropriately modifying $(\widetilde{J}\vert, \widetilde{\Omega}\vert)$ as follows. 
Consider a one-form $K$ on $\mathbb{C}^n$ with the following properties:  

\begin{enumerate}

\item It is (1,0) (with respect to the  
complex structure of $\mathbb{C}^n$) and vertical:
\eq{
P(K)=K
~.}
Starting with a non-vertical $K$, this condition 
can always be satisfied by acting on $K$ with the projector $P$.
\item It is an eigenform of $\mathcal{L}_{\mathrm{Im}V^a}$ (\ie{} 
it has definite  $Q^a$-charge):
\eq{
\mathcal{L}_{\mathrm{Im}V^a}K=q^a K
~,} 
where $q^a$  is half the $Q^a$-charge of ${\Omega}_{\mathbb{C}}$:
\eq{
\mathcal{L}_{\mathrm{Im}V^a}{\Omega}_{\mathbb{C}}=2q^a {\Omega}_{\mathbb{C}}
~.} 
\item It is nowhere-vanishing, and hence can be normalized to:
\eq{
K\cdot K^*=2
~,}
where the dot on the left-hand side denotes contraction 
of indices  with respect to the canonical metric on $\mathbb{C}^n$.

\end{enumerate}

Provided the above conditions are satisfied, the pair $(j,\omega)$ defined by:
\eq{\spl{\label{su2}
\omega&:= -\frac{i}{2}~\! K^*\cdot \widetilde{\Omega}\vert\\
j&:=\widetilde{J}\vert-\frac{i}{2}K\wedge K^*
~,}}
forms a local $SU(2)$ structure obeying
\eq{\spl{\label{su2c}
\omega\wedge\omega^*&= 2j\wedge j\\
\omega\wedge j&=0
~.}}
For further details on $SU(2)$ structures on six-manifolds, see \cite{gauntlett,blt}.

We then define a real two-form $J$ and a complex three-form $\Omega$ by:
\boxedeq{\spl{\label{fsu3}
J&:=\alpha j- \frac{i\beta^2}{2}K\wedge{K}^*\\
\Omega&:=e^{i\gamma}\alpha\beta  K^*\wedge\omega
~,}}
where $\alpha$, $\beta$, $\gamma$ are real gauge-invariant functions subject to 
the condition that $\alpha$, $\beta$ are nowhere-vanishing on $\mathcal{M}_6$, but otherwise arbitrary. 
Provided there exists a one-form $K$ with the properties listed above, 
the pair  $(J,\Omega)$ in (\ref{fsu3}) gives a well-defined family of 
$SU(3)$ structures on $\cm_{6}$. The above construction is a generalization of the one which was used in \cite{toma,gt} to define an $SU(3)$ structure on $\mathbb{CP}^3$.

For $\alpha=\beta=1$, $\gamma=\pi/2$, the metric associated with $(J,\Omega)$ 
is precisely the one induced from the canonical metric on $\mathbb{C}^n$. This 
can be seen by noting that for $\alpha=\beta=1$, $\gamma=\pi/2$ the global $SU(3)$ structure (\ref{fsu3}) is related to the local $SU(3)$ structure $(\widetilde{J}|,\widetilde{\Omega}|)$ constructed from (\ref{jdef}),(\ref{omans}) through $K\leftrightarrow K^*$. Precisely at this point in the parameter space of $SU(3)$ structures $\widetilde{\mathcal{I}}^2$ is strictly negative, as required for complex decomposability -- see the discussion around eqn.~(\ref{hitcncl}). It is reasonable to assume that the property of decomposability will hold more generally in a neighborhood of that point in the parameter space. We would like to stress however that, although we know of no counterexamples, there is no general proof of that statement\footnote{We would like to thank Sheldon Katz for discussions on this point.}. Therefore, away from the $\alpha=\beta=1$, $\gamma=\pi/2$ point, complex decomposability has to be checked explicitly. We have confirmed that the decomposability indeed holds for the examples presented in the following in section \ref{examples}, for generic constant values of $\alpha$, $\beta$, $\gamma$.

The torsion classes 
corresponding to the $SU(3)$ structure   $(J,\Omega)$ (\ref{fsu3}) 
are not specified at this point and must be 
read off of eqs.~(\ref{torsionclass}). Demanding that the torsion 
classes be of a specific type  will impose further restrictions.

In the next section we carry out this procedure explicitly for 
a couple of  specific examples of smooth, compact, three-dimensional toric varieties described in section \ref{toricreview}.

\subsection{Examples}\label{examples}

To illustrate the procedure for constructing 
$SU(3)$ structures on SCTV explained in section \ref{sec:4.2}, we will now consider a couple of explicit examples. The first 
one is the case $\mathcal{M}_6=\mathbb{CP}^3$. The construction of 
$SU(3)$ structures on $\mathbb{CP}^3$ has been worked out before, both by regarding $\mathbb{CP}^3$ as a twistor space \cite{toma} and as a coset \cite{klt}, as well as in the special 
nearly-K\"{a}hler limit \cite{bc}. As already mentioned, our present  approach is a generalization of \cite{toma,gt}, although we do not make use of the twistor description here. 

The second example is new: it is the toric $\mathbb{CP}^1$ bundle over the 
(infinite) class of two-dimensional toric varieties with $n=4$ listed at the end of  section \ref{symplectic}.

\subsection*{Example 1: $\mathbb{CP}^3$}

There is a single charge in this example, given by:
\eq{
\label{eq:cha}
Q^1=(1,1,1,1)~.
}
The vertical one-form $K$ can be taken to be:
\eq{\label{kcp3}
K=\left( \sum_{i=1}^4|z_i|^2 \right)^{-1/2}(-z^2dz^1+z^1dz^2-z^4dz^3+z^3dz^4)
~,}
which, as is straightforward to verify, satisfies all conditions 
listed in section \ref{sec:4.2}. The $SU(3)$ structure is then given 
by (\ref{fsu3}).

The torsion classes of the $SU(3)$ structure can be straightforwardly 
computed. The only non-vanishing ones are $\mathcal{W}_{1,2}$ \cite{klt, toma}; hence this example is of the type of \cite{lt}.

\subsection*{Example 2: Toric $\mathbb{CP}^1$ bundle}

Here we study the $\mathbb{CP}^1$ bundle over the infinite class of  two-dimensional toric varieties with $n=4$. As described at the 
end of section \ref{symplectic}, the charges for this fibration are
\eq{
\label{eq:charges}
Q^1=(0,1,0,1,c_1,0) \quad Q^2=(1,a,1,0,c_2,0) \quad Q^3=(0,0,0,0,1,1)~,
}
where $a$, $c_1$, $c_2\in\mathbb{Z}$. In the following we will assume $a\leq 0$. 
Moreover, as we will show below, $c_1$, $c_2$ will be restricted 
to the values given in eq.~(\ref{rv}).

Requiring the manifold to be non-singular restricts the range of the K\"ahler moduli $\xi^a$ so that $\widetilde{J}|$ lies inside the K\"ahler cone, whose boundaries we now derive. There is a set of two-cycles $C^a$, $a=1,2,3$,  associated with the charges $Q_i^a$, defined via
\eq{
D_i\cdot C^a=Q^a_i
~,}
where $D_i$, $i=1,\dots 6$ are the holomorphic divisors given by $z_i=0$. We would like to choose the generators of the gauge group $U(1)^3$ so that the $C^a$'s are generators of the Mori cone (i.e. the set of all two-cycle classes with holomorphic representatives).  

If we make a change of basis so that:
\eq{
\widetilde{Q}^1=Q^1~;~~\widetilde{Q}^2=Q^2-c_2Q^3;~~\widetilde{Q}^3=Q^3~,
}
we find that in this new basis the moment maps read
\eq{\spl{
\label{mmaps}
|z^2|^2+|z^4|^2-2|z^5|^2&=\widetilde{\xi}^1\\
|z^1|^2+a|z^2|^2+|z^3|^2+(a-2)|z^6|^2&=\widetilde{\xi}^2\\
|z^5|^2+|z^6|^2&=\widetilde{\xi}^3~,
}}
where we have taken (\ref{rv}) into account. Furthermore, 
the K\"ahler cone is given by 
$\widetilde{\xi}^a>0$, as we now show.

There are exactly three linearly independent divisors, which can be taken to be $D_1$, $D_2$, $D_5$, since the gauge invariance implies:
\eq{
D_3=D_1~;~~D_4=-a~\! D_1+D_2~;~~D_6=-(2+a)D_1+2D_2+D_5~.
}
Taking the above equivalences into account, it can be seen that  the two-cycles $\widetilde{C}^a$  are given by
\eq{
\widetilde{C}^1=D_1\cdot D_5~;~~\widetilde{C}^2=(a-2)D_1\cdot D_2+D_2\cdot D_5~; ~~\widetilde{C}^3=D_1\cdot D_2~.
}
Using the above relations to express the holomorphic curves (transversal intersections) $D_i\cdot D_j$ in terms of the $\widetilde{C}^a$'s, one can see that the latter generate (with positive coefficients) the full Mori cone. It then follows that the K\"{a}hler cone is given by  $\widetilde{\xi}^a>0$, as promised.

We would like to find a one-form $K$ that obeys the
three conditions listed in section \ref{sec:4.2}. Let us consider the following ansatz for $K$:
\eq{\label{eq:K}
K= \alpha_1 K_1 + \alpha_2 K_2~,
}
where
\eq{\spl{
K_1& :=- z^3 dz^1+ z^1 dz^3~, \\
K_2& := -z^4dz^2+z^2dz^4+ \frac{a  z^2 z^4}{|z^1|^2+|z^3|^2} \left(\bar{z_1}dz^1 +\bar{z_3}dz^3 \right)~,
}}
and $\alpha_i$ are functions of $z^i$ that we will determine in the following. 
It can be seen that the $K_1$, $K_2$ defined above are mutually orthogonal eigenvectors of $P_{ij}$ with eigenvalue equal to one, and therefore 
this ansatz guarantees that the first condition of section \ref{sec:4.2} is satisfied. Moreover we have:
\eq{\spl{
|K_1|^2&=2\left( |z^1|^2+|z^3|^2\right)\\
|K_2|^2&=2\left( |z^2|^2\left(1+ \frac{a^2|z^4|^2}{|z^1|^2+|z^3|^2} \right)+ |z^4|^2 \right)
~,}}
and hence $|K_1|^2$, $|K_2|^2$ are strictly positive, as follows from 
the moment maps (\ref{mmaps}) and the fact that $a\leq 0$.

Taking the mutual orthogonality of $K_1$, $K_2$ into account, we see that the following choice for $\alpha_i$
\eq{\label{427}
\alpha_1=z^6 \quad \alpha_2=z^5~,
}
would guarantee that $|K|^2$ is strictly positive, and hence can be normalized in accordance with the third condition of section \ref{sec:4.2}.  Indeed, it follows from the 
above that
\eq{
|K|^2 = |z^6|^2 |K_1|^2 + |z^5|^2 |K_2|^2~
}
is strictly positive since 
$|K_1|^2$, $|K_2|^2$ are both strictly positive and $z_5$, $z_6$ cannot both be zero, as follows from the last moment map in (\ref{mmaps}).

Finally, it is straightforward to see that the second condition  of section \ref{sec:4.2} is also satisfied, provided we set
\eq{\label{rv}
c_1=-2~;~~~~~c_2=2-a~.
}
Therefore the one-form $K$ constructed above obeys all the conditions 
of section \ref{sec:4.2} and gives rise to a well-defined $SU(3)$ structure 
via (\ref{fsu3}).  Note that 
the choice in eqn.~(\ref{427}) can be easily generalized by multiplying $\alpha_i$ with arbitrary nowhere-vanishing gauge-invariant functions.

Using the formul\ae{} of section \ref{symplectic}, it is straightforward to read off the associated torsion classes from the exterior derivatives of $J$ and $\Omega$. 
The explicit expressions for $a=0$, constant $\alpha$, $\beta$ and $\gamma=\pi/2$  
are given in appendix \ref{appendixb}.

Let us finally mention that upon 
gauge-fixing the $U(1)^3$ gauge freedom of the symplectic description, one could easily solve the moment maps (\ref{mmaps}) explicitly,  thereby obtaining a description of the $SU(3)$ structure of $\mathcal{M}_6$ in terms of six real coordinates.

\section{Conclusions}\label{conclusions}

Toric varieties have so far  played an intermediate role in string compactifications: one uses them either 
as auxiliary spaces in which  Calabi-Yau manifolds are embedded, 
or as local (usually singular and noncompact) descriptions of the geometry. 
The present paper proposes a paradigm shift: once fluxes are turned on we may use the three-dimensional  (six real dimensions) 
smooth, compact toric varieties themselves as compactification manifolds. 

For this construction to work one needs to equip the SCTV with a so-called `$G$-structure', which is a reduction of the structure 
group of the manifold. In the case of $SU(3)$ structures in particular, supersymmetry `selects' an almost complex structure 
which in general may not be integrable. The point, however, is that due to 
the presence of a  second, integrable complex structure, one can still use the underlying algebraic-geometrical (toric) description 
of the manifold in order to work out all the relevant objects relating to the $G$-structure. Indeed this was the 
main motivation of the present paper: to bring the powerful tools of algebraic geometry to bear on the subject of 
flux compactifications and $G$-structures. 

There are by now extensive databases of CY threefolds which  can be scanned systematically in search
 for models with realistic phenomenology. It is our hope that the present paper is a step towards 
an analogous systematization for compactifications with non-vanishing flux -- in which 
case $\mathcal{M}_6$ is no longer a CY. For example, although for the purposes of the present paper 
we have limited ourselves to the first few examples of SCTV, the simple $SU(3)$ ansatz of section \ref{sec:4.2} could  
be straightforwardly implemented in a systematic computerized scan of three-dimensional SCTV. 

Although our main focus has been three-dimensional SCTV, the tools developed in the present paper ({\it e.g.} for reading 
off the torsion classes of a given $G$-structure) are rather general. In particular, it would be interesting to apply our methods to  
smooth, compact toric varieties of other dimensionalities, as well as different $G$-structure ans\"{a}tze than 
the one used in section \ref{sec:4.2}. Other possible extensions of the present work 
include noncompact ({\it e.g.} conical) toric manifolds, or manifolds of the form $\mathcal{M}_4\times T^2$, where 
 $\mathcal{M}_4$ is a two-dimensional smooth, compact toric variety. 

Furthermore, our work opens up the possibility  for several applications. It is plausible that among the SCTV discussed in this paper, there may be candidates for 
compactification to classical (meta-) stable dS vacua (see \cite{Cav, Dan} for some recent work and further references).  
This is a long-standing problem in flux compactifications, and addressing it would put string-inspired cosmology under better control. 
Other applications include supersymmetric compactifications on SCTV to ${\mathbb R}^{3,1}$ space, heterotic compactifications\footnote{Having a SCTV 
as an internal manifold in heterotic compactifications may facilitate the construction of the gauge bundle.}, and non-supersymmetric vacua along the lines of \cite{nonsusy1,nonsusy2}.

It is also plausible that several new AdS vacua can be constructed on SCTV, e.g. using the scalar ansatz of \cite{Lust:2009zb,Lust:2009mb}, which could be gravity duals of gauge theories. It would be very interesting to see how much of the already available toric `technology' for reading off the dual gauge theory can be carried over to the case where the gravity dual is based on a SCTV. We hope to 
return to this in the future.

\appendix

\section{Torsion classes for Example 2: Toric $\mathbb{CP}^1$ bundle}\label{appendixb}

The torsion classes of the toric $\mathbb{CP}^1$ bundle (the second 
example of section 
\ref{examples}) can be straightforwardly computed using the 
formul\ae{} of section \ref{symplectic}. In the patch where $z_1, z_4, z_6$ are all non-zero, for constant $\alpha$, $\beta$, $\gamma=\pi/2$ in eqs.~(\ref{fsu3}), and specializing to the case $a=0$ for simplicity, we find:
\begin{align}\label{torsions}
&\mathcal{W}_1=- \frac{2 i}{3\sqrt{p_1}p_2} \frac{ (\alpha+\beta^2)}{ \alpha \beta} \nn\\
&\mathcal{W}_3 = - \frac{ (\alpha+\beta^2)}{\beta^2} \Psi \wedge \left(\alpha j + i \frac{\beta^2}{2} K \wedge K^* \right)\nn\\
&\mathcal{W}_4 =  \frac{ (\alpha+\beta^2)}{\beta^2} \Psi~,
\end{align}
where
\begin{align}
&p_1 := \left( |z^5|^2 + |z^6|^2\right) \left(|z^2|^2+|z^4|^2 \right)\left(|z^1|^2+|z^3|^2 \right) +
4|z^5 z^6|^2 \sum_{i=1}^{4} |z^i|^2, 
\nn\\
&p_2 := |z^5|^2 \left(|z^2|^2+|z^4|^2 \right)+
|z^6|^2 \left(|z^1|^2+|z^3|^2 \right) 
\end{align}
are real, gauge-invariant, nowhere-vanishing functions, and
\eq{
\Psi :=\frac{1}{p_2} \left( |z^5|^2 -|z^6|^2  \right)\mathrm{Re}(\bar{z}_5 \mathcal{D}z^5)}
is a real one-form. It is straightforward to check that $\mathcal{W}_3,\mathcal{W}_4$ are primitive. The torsion classes 
$\mathcal{W}_2,\mathcal{W}_5$ are nonzero, but their explicit expressions are 
lengthy and not particularly illuminating, so we do not list them here. 
As can be seen from (\ref{torsions}), 
in the case $\alpha=-\beta^2$ the only non-vanishing torsion classes are $\mathcal{W}_2$, $\mathcal{W}_5$. Hence this example does {\it not} lead to  a IIA solution of the form of \cite{lt}, even in the massless limit. 

In deriving the various properties of the $SU(3)$ structure, we have made use of the symbolic computer program \cite{edc}. We have also  verified that the 
Hitchin functional is strictly negative, as it should.

%
%

\end{document}